\documentclass{nature}
\usepackage{times}

\usepackage[english]{babel}
\usepackage{lipsum}  
\usepackage{amsmath}
\usepackage{graphicx}
\usepackage{svg}
\usepackage{tabularx,multirow}
\usepackage{longtable,threeparttablex}
\graphicspath{{figs/}}
\usepackage{hyperref}
\hypersetup{
 bookmarks=true,		
 unicode=false,			
 pdftoolbar=true,		
 pdffitwindow=false,		
 pdfstartview={FitH},		
 pdfcreator={pdflatex},		
 pdfnewwindow=true,		
 colorlinks=true,		
 linktoc=page,			
 linkcolor=blue,		
 citecolor=blue,		
 filecolor=blue,		
 urlcolor=blue			
}

\begin{document}

\title{Carbon-contaminated topological defects in hexagonal boron nitride for quantum photonics }

\author{Rohit Babar$^{1,2,*}$, \'{A}d\'{a}m Ganyecz$^{1,2}$, Igor A. Abrikosov$^{3}$, Gergely Barcza$^{1,2}$, Viktor Iv\'{a}dy$^{2,3,4,*}$} 

\maketitle

\begin{affiliations}
\item  {Wigner Research Centre for Physics, PO Box 49, H-1525, Budapest, Hungary}
\item {MTA-ELTE Lend\"{u}let "Momentum" NewQubit Research Group, P\'{a}zm\'{a}ny P\'{e}ter, S\'{e}t\'{a}ny 1/A, 1117 Budapest, Hungary}
\item   {Department of Physics, Chemistry and Biology, Link\"oping University, SE-581 83 Link\"oping, Sweden}
\item{Department of Physics of Complex Systems, E\"otv\"os Lor\'{a}nd University, Egyetem t\'{e}r 1-3, H-1053 Budapest, Hungary}
\item[*] email: {rohit.babar@wigner.hun-ren.hu, ivady.viktor@ttk.elte.hu}
\end{affiliations}

\date{\today}

\vspace{1cm}

\begin{abstract}
Topological defects, such as Stone-Wales defects and grain boundaries, are common in 2D materials. In this study, we investigate the intricate interplay of topological defects and carbon contamination in hexagonal boron nitride revealing an intriguing class of color centers. We demonstrate that both carbon contamination and strain can stabilize Stone-Wales configurations and give rise to emitters with desirable optical properties in the visible spectral range. Inspired by these results, we further demonstrate that carbon atoms at grain boundaries can resolve energetic B-B and N-N bonds leading to highly favorable atomic structures that may facilitate the accumulation of carbon contamination at the boundaries. Similarly to contaminated Stone-Wales defects, carbon-doped grain boundaries can also give rise to color centers emitting in the visible spectral range with short radiative lifetime and high Debye-Waller factors.
Our discoveries shed light on an exciting class of defects and pave the way toward the identification of color centers and single photon emitters in hBN.
\end{abstract}

\section{Introduction}

Wide-bandgap layered semiconductors have recently emerged as a capable platform for developing integrated photonic devices.\cite{Liu_2022,Li2023} Similar to conventional bulk semiconductors, atomic-scale structural defects in these materials can implement single photon sources with desirable optical properties.\cite{Vasconcellos_2022} While layered materials may contain defects and impurities in high concentrations, exfoliation of the layered structure provides an ultimate method to control the number of defects in the sample. Thin flakes may contain emitters in such a low number that individual emitters can be resolved by confocal microscopy.\cite{Tran2015} Furthermore, due to the absence of internal reflection in thin samples, photons emitted by the color centers can be collected with near unity efficiency.\cite{Schell2018,Nikolay2019,Boll2020} These attributes make layered semiconductors a desirable platform for integrated photonics.

Hexagonal boron nitride is famous for its applicable single photon sources\cite{caldwell_photonics_2019, sajid_single-photon_2020} emitting in a broad spectral range starting at the near-infrared region\cite{camphausen_observation_2020}, involving the visible spectrum\cite{tran_robust_2016,chejanovsky_single-spin_2021}, and finishing at the UV-B ultraviolet spectral region\cite{bourrellier_bright_2016}. It is generally accepted today that these emitters are related to structural defects of the lattice forming deep levels in the 6.1~eV band gap of hBN. The emitters that have been reported in the literature possess a wide variety of properties, some of them exhibit a bright emission in the zero-phonon line (ZPL)\cite{caldwell_photonics_2019}, while others emit in a broad spectrum. It has been recently demonstrated that carbon contamination plays a crucial role in the fabrication of visible quantum emitters.\cite{Mendelson2019,kumar_localized_2023,zhong_carbon-related_2024}

Despite the numerous experimental and theoretical studies\cite{sajid_single-photon_2020}, the microscopic structure of most of the emitters is yet to be revealed. A plethora of native\cite{ivady_ab_2020,weston_native_2018,hamdi_stonewales_2020} and impurity-related defects\cite{mackoit-sinkeviciene_carbon_2019,PhysRevB.105.184101,C7NR04270A,Sajid2018,Auburger2021,Jara2021,Cholsuk2022,bhang_first-principles_2021,Li2022} has been computationally studied in an attempt to account for the experiential observations. Except for a few  examples\cite{ivady_ab_2020,Li2022}, no consensus has been achieved so far regarding the microscopic origin of the emitters. This raises the question of whether all the relevant defect structures have been taken into consideration.

It is known from studies on graphene that the formation of topological defects, such as the Stone-Wales (SW) defects\cite{tiwari_stonewales_2023} and grain boundaries\cite{Yazyev2014} in the honeycomb lattice of graphene is common. The formation of SW defect in hBN is, however, less favorable, due to the polarized nature of the B-N bonds.\cite{Charlier1999} Indeed, the formation energy of the 5-7 ring SW defect is as high as 7.2~eV \cite{Li2008,Chen2009,Wang2016} in hBN compared to the 5.1~eV of graphene SW defect \cite{PhysRevLett.100.175503}. These results suggest that topological SW defects are less common in hBN. It is worth mentioning that low-energy irradiation may provide sufficient energy to the lattice to form various SW defects, see Ref.~[\citen{Gibb2013}].

The formation of grain boundaries, featuring various types of topological defects, is governed not only by thermodynamics but also by the conditions of the growth. In particular, simultaneous nucleation of hBN seeds on the substrate may lead to hBN grains of different orientations, reversed edges, and various relative shifts deviating from the periodicity of the hBN lattice. The interfaces of such grains give rise to various grain boundaries, which form topological line defects~\cite{Auwrter2003,PhysRevLett.104.096102}. Grain boundaries in hBN have been observed by transmission electron (TEM), scanning tunneling (STM) and atomic force microscopy (AFM)\cite{Gibb2013,Cretu2014,Li2015,Bayer2017,Ren2019NL,PhysRevMaterials.3.014004,Park2020,Wrigley2021} and studied computationally by density functional theory (DFT).\cite{Liu2012,Li2012,Ding2014} Recently, quantum emitters from grain boundaries in hBN have been reported.\cite{Chejanovsky2016,NgocMyDuong2018,Xu2018,Vogl2019,Mendelson2019,Shevitski_2019} Bright emitters along lines in hBN have been speculated to be related to SW defects.\cite{Chejanovsky2016}  Furthermore, cathodoluminescence (CL) study reveals 5.46 and 5.63 eV emission lines (D series) localized at extended defects in hBN.\cite{PhysRevB.79.193104,Jaffrennou2007} Their origin is attributed to excitons bound to grain boundary, dislocations \cite{Jaffrennou2007,PhysRevB.89.035414} or stacking faults\cite{Watanabe2006a,Bourrellier2014}. Recently, the presence of PL and CL emission around 2.3 eV is proposed to originate from an array of dislocations.\cite{Ciampalini2022}

In this article, we show that the formation energy of topological defects can be reduced by incorporating carbon contamination into the structure. To demonstrate this, we study a set of carbon-containing Stone-Wales defects (SW-C), with 5-7 rings and 1-4 carbon contaminants in different charge states. In addition, we study  10 different types of grain boundaries and numerous carbon dimer configurations adjacent to them. We show that the formation energy of SW-C defects and carbon dimers at certain positions of the grain boundary is significantly reduced. We also show that applying tensile strain to the hBN sample further favors the formation of SW-C defect. Furthermore, we study the energy balance of different grain boundary-adatom configurations and conclude that mobile carbon atoms may accumulate at the grain boundaries and form stable carbon-contaminated line defects. Finally, we show that the SW-C defects as well as the favorable carbon dimers at the grain boundary can give rise to color centers with short excited state lifetime and high Debye-Waller factor in the visible spectral region. The optical properties of the studied defect structures resemble the signatures of the not yet identified carbon-related defects in hBN. 


\vspace{1cm}

\section{Results and discussions}

\subsection{Formation energy of carbon-containing Stone-Wales defects in hBN}

\begin{figure*}
\includegraphics[width=1.0\textwidth]{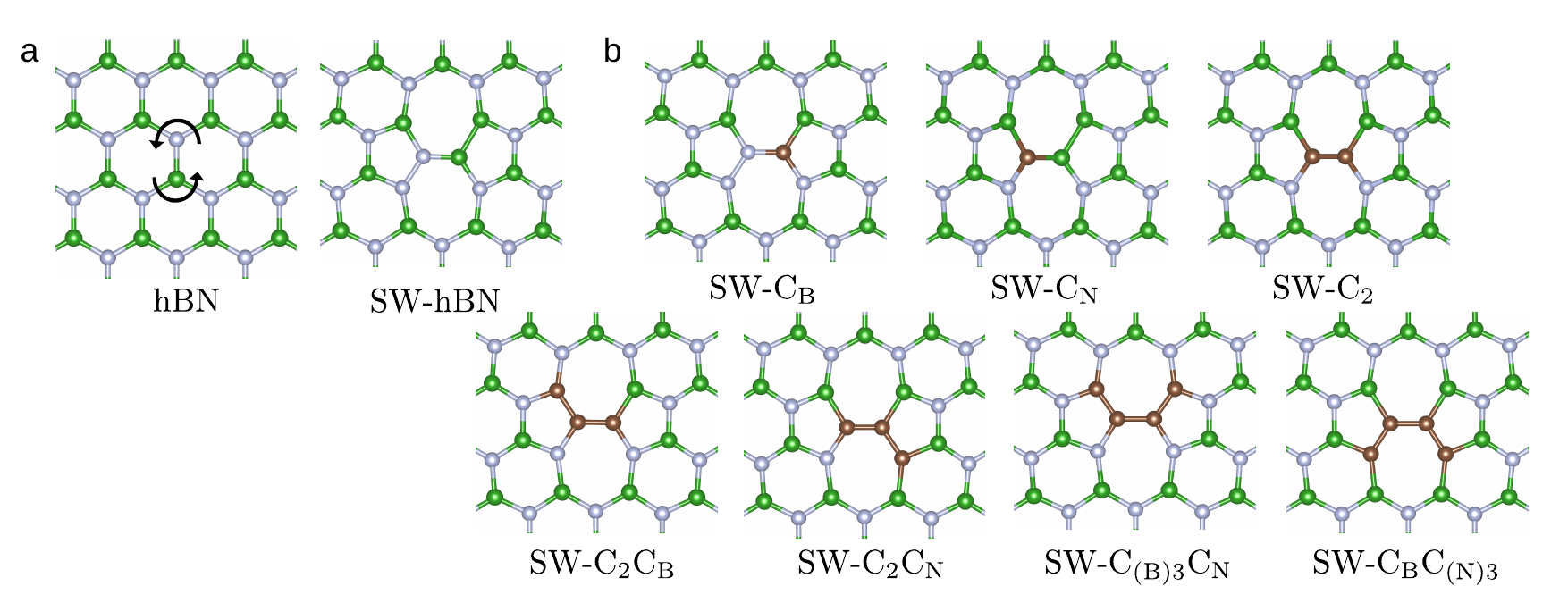}
	\caption{Stone-Wales defect in pure and carbon doped hBN. (a) The ideal honeycomb lattice of hBN transforms to SW defect through a bond rotation indicated by the curved arrows. (b) SW-C defect counterparts for the most prevalent carbon clusters in hBN. Green, gray, and brown spheres denote boron, nitrogen, and carbon atoms respectively.}
	\label{fig:fig_struc}  
\end{figure*}

In Fig.~\ref{fig:fig_struc} we depict the considered Stone-Wales and carbon-containing Stone-Wales configurations. In a pristine hBN layer, the 5-7 SW defect can be obtained by "rotating" a first neighbor B-N pair by 90$^\circ$, see Fig.~\ref{fig:fig_struc}(a). The numbers 5 and 7 indicate the number of vertices of polygons formed by the SW defect. In addition to the 5-7 SW defect,  we investigate various SW-C structures obtained from carbon substitutional defects by rotating either a C-N, a C-B, or a C-C bond by 90$^\circ$. In Fig.~\ref{fig:fig_struc}(b), we depict the resulting configurations including either one, two, three, or four carbon atoms. We note that not all the possible configurations are depicted in Fig.~\ref{fig:fig_struc}.  We focus on those configurations where the carbon atoms are connected and centered around the shared atoms of the 5 and 7 rings, see Fig.~\ref{fig:fig_struc}. In general, we find these configurations to be energetically the most favorable ones, see related results later and in the Supplementary Information. Furthermore, we note that the SW-C$_{\text{B}}$, SW-C$_{\text{N}}$, and SW-C$_{\text{2}}$ configurations tend to buckle the 2D hBN flake at the defect site. Such out-of-plane relaxations are not observed for other SW configurations containing more than two carbon atoms.

Next, we compute $F_{\text{SW-C-cluster}}^{\text{rel}}$, the \emph{relative} formation energy of the SW-C defects compared to the formation energy of the corresponding carbon substitutional defect as
\begin{equation} \label{eq:dif}
    \begin{split}
        F_{\text{SW-C-cluster}}^{\text{rel}} \left( q \right) = 
        \Delta E_{\text{SW-C-cluster}}^{\text{tot}} \left( q \right) = \\
        E^{\text{tot}}_{\text{SW-C-cluster}} \left( q \right) - E^{\text{tot}}_{\text{C-cluster}} \left( q \right)
    \end{split}
\end{equation}
where $E^{\text{tot}}_{\text{SW-C-cluster}}\left( q \right)$ and $E^{\text{tot}}_{\text{C-cluster}}\left( q \right)$ are the total energies of defective supercells in charge state $q$.  $F_{\text{SW-C-cluster}}^{\text{rel}} \left( q \right)$ measures the energy cost of transforming an existing carbon impurity (cluster) into a related SW-C configurations. Since carbon contamination can be found in hBN in high concentrations, here we discard the energy cost of forming carbon substitutional defects and clusters. The total energy difference can be equal to the difference of the formation energy~\cite{Gali2019} of the defects assuming the following requirements are fulfilled: (a) the structures contain the same number of atoms, (b) the finite size corrections for the total energy of the SW-C-cluster and the C-cluster are approximately equal and thus they cancel out, and (c) there exists a Fermi energy interval in the band gap of hBN, where both defects possess the same charge state. 
Condition a) is trivially fulfilled in our simulations, while condition b) should be ensured by the large models considered in our calculations. Condition c) is expected to be fulfilled when the Fermi energy is in the middle of the bandgap, see discussions of the electronic structure later. Therefore, we compute the relative formation energy of the SW-C defects according to Eq.~(\ref{eq:dif}).

\begin{figure}
\begin{center}
\includegraphics[width=0.5\textwidth]{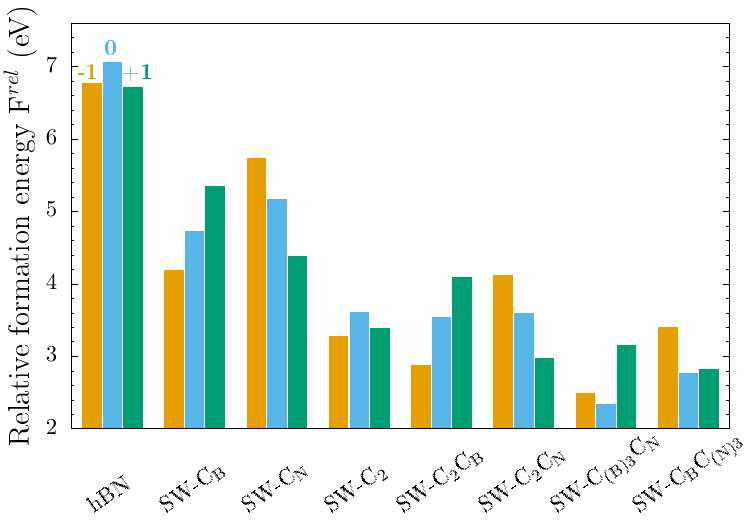}
\end{center}
	\caption{Formation energy differences of SW-C defects and corresponding carbon substitutions complexes for different charge states, negative (amber columns), neutral (light blue columns), and positive (green columns).}
	\label{fig:fig_energy}  
\end{figure}
As shown in Fig.~\ref{fig:fig_energy}, the relative formation energy of planar SW-C structures reduces significantly by replacing 1, 2, 3, or 4 boron and nitrogen atoms of the SW defect. In the neutral charge state, replacing one boron (nitrogen) with a carbon atom results in a 33~\% (27~\%) reduction of the relative formation energy of the SW configurations.  Replacing the boron and nitrogen atoms in between the 5 and 7 atoms rings results in an even larger, 49~\% reduction of the relative formation energy of the SW configurations. Including three carbon atoms in the SW configuration does not decrease the formation energy further compared to the SW-C$_2$ configuration. Finally, including 4 carbon atoms in the SW structure considerably lowers the energy differences compared to the SW-C$_2$ configuration.

To explain our observations, we start with a discussion of the stability of the SW configuration. In the case of the hBN-SW defect, B-B and N-N bonds are formed, which are energetically unfavorable. Incorporating carbon impurities in the structure relaxes the frustrated bonds through the replacement of B-B and N-N bonds with C-N and C-B bonds. The energy gain for replacing the B and N atoms shared by the 5 and 7 rings is 2.3~eV and 1.9~eV, respectively. The carbon substitution at the next neighbor site that resolves only one-half of the frustrated bond is less favorable in case of SW-C$_{\text{B}}$ while such an arrangement is preferred for SW-C$_{\text{N}}$. When both middle atoms are replaced by carbon, the energy gain is approximately the sum of these two contributions, i.e.\  3.6~eV. Two carbon atoms can resolve both energetically unfavorable bonds of the hBN SW defect. The inclusion of an additional carbon atom does not lower the formation energy further, see Fig.~\ref{fig:fig_energy}.
The inclusion of four carbon atoms in the SW defect reduces the formation energy further due to increased replacement of C-N and C-B bonds with C-C bonds.\cite{PhysRevB.73.073108,marek_thermodynamics_2022}  The inclusion of disjoint clusters of carbon atoms at the periphery of the SW defect does not significantly affect the formation energy, thus the configurations depicted in Fig.~\ref{fig:fig_struc} can be considered the most favorable ones for a given stoichiometry. 

Considering the charge state dependence of the relative formation energy of the SW-C defects, we observe a clear pattern for an odd number of carbon atoms. When more boron atoms are substituted by carbon than nitrogen, i.e.\ SW-C$_{\text{B}}$ and SW-C$_{\text{2}}$C$_{\text{B}}$, the negative charge further stabilizes the SW configuration, while the positive charges increase the energy of the SW configuration, see Fig.~\ref{fig:fig_energy}. The opposite is true for the SW-C defects where more N atoms are replaced by carbon atoms than boron atoms. For an even number of carbon atoms, no clear relationship between the charge state and the relative formation energy is observed.

\begin{figure}
\begin{center}
    \includegraphics[width=0.5\textwidth]{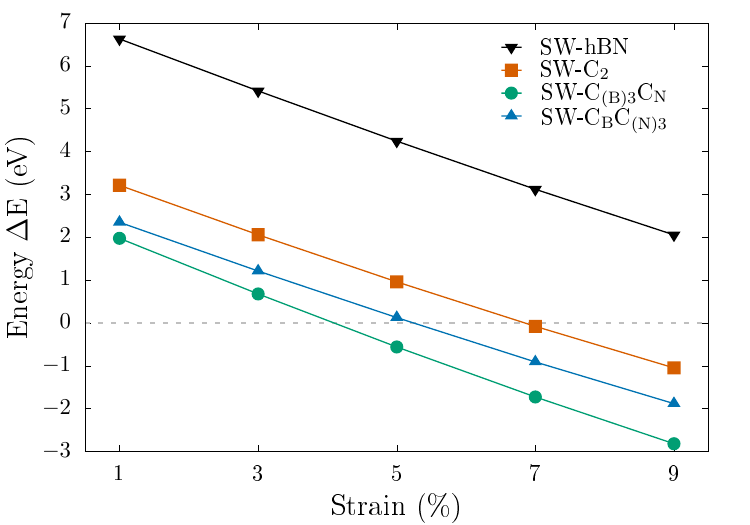}
\end{center}
	\caption{Relative formation energies for SW-C defects under tensile strain. The SW-C defects are favorable for 4-7~\% strain, applied parallel to the SW bond. Here, $\Delta$E refers to the energy difference between the substituted carbon cluster and corresponding SW-C configurations. A negative value denotes the favorable formation of SW-C defects.}
	\label{fig:fig_strain}  
\end{figure}
The SW configuration introduces highly anisotropic strain in the hBN lattice which can help to relax external uniaxial stress. Consequently, the formation of SW configurations is favorable under tension.\cite{tiwari_stonewales_2023} Therefore, we investigate the effect of strain on the formation energy of SW, SW-C$_2$, SW-C$_{\text{(B)3}}$C$_{\text{N}}$, and SW-C$_{\text{B}}$C$_{\text{(N)3}}$ configurations. As can be seen in Fig.~\ref{fig:fig_strain}, a strain applied along the axis passing through the shared atoms between the 5 and 7-atom rings further reduces the relative formation energy of the SW-C defects. Due to the lower relative formation energy of the SW-C defects than the hBN SW defect, the SW-C defect may become more favorable than the corresponding non-topological carbon cluster for an applied strain as large as 4-7\%. We observe that the lower the relative formation energy of the SW-C defects the lower the strain needed to make the SW-C configuration favorable. Therefore, we anticipate that defected regions in hBN with a high concentration of carbon atoms may host Stone-Wales structures.

\subsection{Formation energy of carbon dimers at grain boundaries}

\begin{figure*}
\begin{center}
\includegraphics[width=0.95\textwidth]{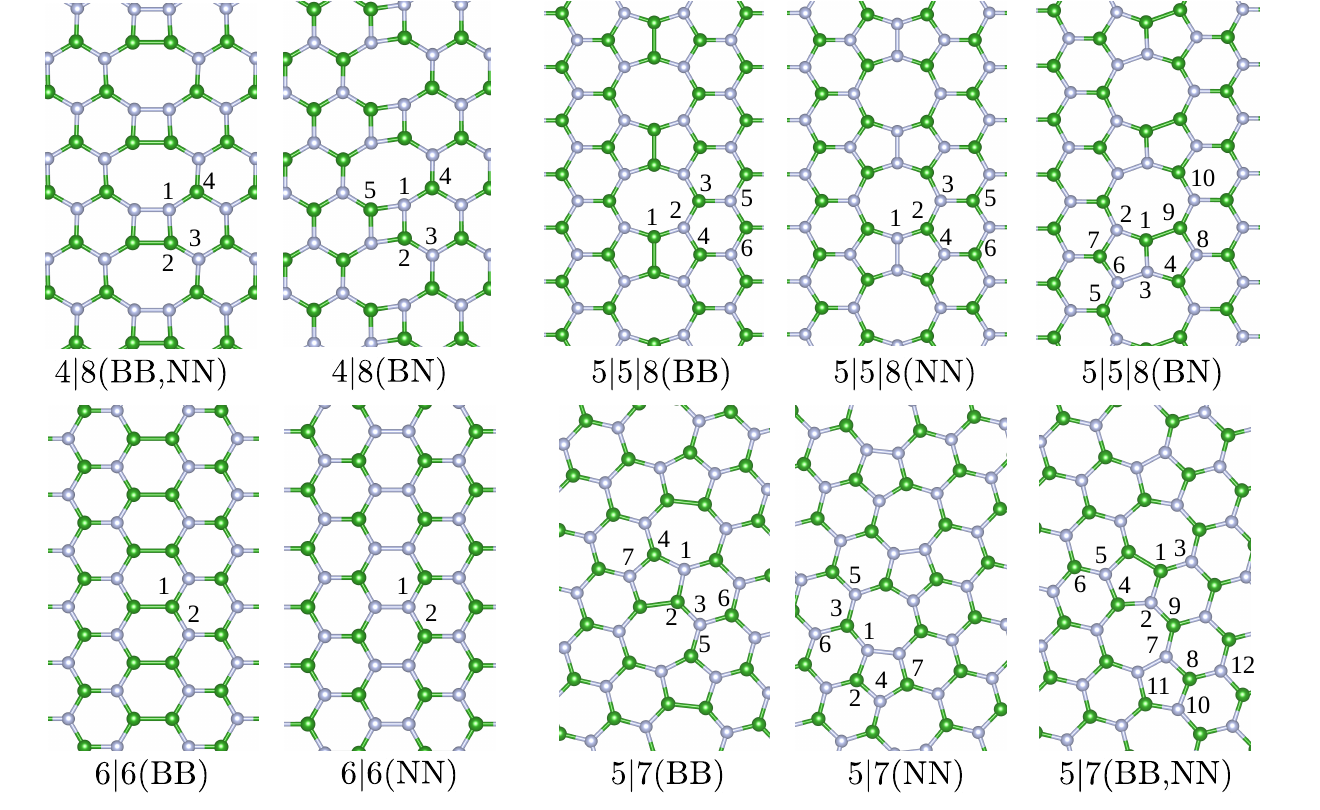}
\vspace{0.5cm}
\includegraphics[width=0.95\textwidth]{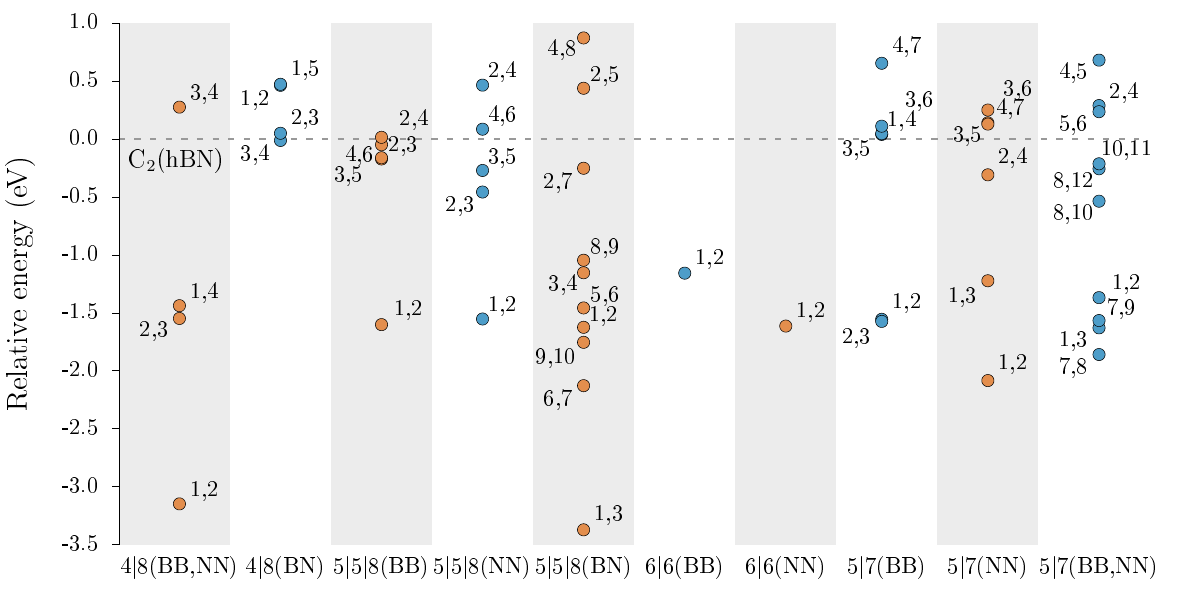}
\end{center}
	\caption{(a) Grain boundaries and unique sites for carbon substitution considered in our study. (b) Favorable sites for carbon dimer at the grain boundary. Amber and blue  dots show the formation energy of carbon dimers at the numbered sites close to the grain boundary. The gray dashed line denotes the reference energy for the carbon dimer away from the respective grain boundary. }
	\label{fig:fig_GB}  
\end{figure*}

The formation of isolated carbon-containing Stone-Wales defects may still be hindered by the relatively low yet positive formation energy of the structures. However, the growth of hBN has the potential to give rise to grain boundaries, inevitably leading to the formation of Stone-Wales-like defects. In hBN, such grain boundaries comprise of line defects that incorporate different types of B-N rings, most often 4,5,6,7, and 8-atom rings, in a certain sequence. Details of the line defect formed depend on the relative orientation and position of the hBN grains interfacing at the grain boundary. In this section, we investigate the potential of carbon contamination to stabilize high energy bonds at grain boundaries, similar to the case of isolated Stone-Wales defects. 

We consider 10 different boundaries as depicted in Fig.~\ref{fig:fig_GB}(a). Most of these grain boundaries have been observed experimentally either in isolated or combined form \cite{Gibb2013,Cretu2014,Li2015,Ren2019NL,Park2020} and studied computationally \cite{Liu2012,Li2012,Gomes2013}. To distinguish different grain boundaries, we introduce the following notation: integers $i$, $j$, and $k$ separated by vertical lines, such as $i|j|k$, define the number of atoms forming the B-N rings at the grain boundary, while letter pairs in parentheses characterize the bonding (B-B, B-N, and N-N) at the interface. The examined grain boundaries can be categorized into four distinct groups based on the sequence of the different B-N rings, i.e.,  there are two $4|8$, three $5|5|8$, two $6|6$, and three $5|7$ type grain boundaries. To accommodate these line defects in periodic supercells, we use a set of single-layer periodic models and nano-ribbons terminated with hydrogen atoms at the edges, see Methods for details. It is worth mentioning that the formation energy of the different types of grain boundaries can differ significantly, however, their appearance is governed not only by thermodynamics but also by the nucleation and growth of the BN nano-flakes. Earlier theoretical studies calculate the grain boundary formation energies between 0.2-1.4~eV/\AA~ depending on the bonding and chemical potential.\cite{Liu2012,Gomes2013,Ding2014,Ren2019NL} High formation energy boundaries can also be observed in experiments.\cite{Ren2019NL} 

In the following, we consider one type of carbon defect at the grain boundaries, namely the highly stable carbon dimer.\cite{mackoit-sinkeviciene_carbon_2019,marek_thermodynamics_2022} Within each grain boundary type, we identify a non-exhaustive set of atomic sites close to the boundary and number them from 1 to 12, see Fig.~\ref{fig:fig_GB}(a). In addition, we consider an adjacent B-N pair far away from the boundary, referred to as far site hereinafter (not shown in Fig.~\ref{fig:fig_GB}(a)). We introduce carbon dimers by replacing adjacent B and N atoms at the numbered sites as well as at the far site. The carbon dimers are referred to as C$_x$-C$_y$, where $x$ and $y$ site indexes are specified in Fig.~\ref{fig:fig_GB}(a). B-B and N-N first neighbor pairs are not considered in our study to avoid dependence of the relative formation energy on the chemical potential of boron, carbon, and nitrogen, which may overwhelm the trends we are interested in. This way we directly compare the formation energy of the near-site dimers to the far-site dimer. The far-site carbon dimer is at least 12~\AA~away from the grain boundary, where the dimer experiences a bulk-like environment. Due to the large number of models and defect configurations considered, we use the computationally affordable PBE\cite{PBE} exchange-correlation functional in the study of the grain boundaries. 

Fig.~\ref{fig:fig_GB}(b) depicts the relative formation energy of various carbon dimers at different types of grain boundaries compared to the formation energy of the far-site carbon dimer, which serves as the reference for the energy scale. Negative (positive) values mean favorable (unfavorable) configurations compared to the far-site carbon dimer. Our calculations identify several favorable carbon dimer sites at the grain boundaries. Taking a closer look, we can observe a similar trend as for the isolated 5-7 Stone-Wales defects, i.e.\ replacement of unfavorable B-B and N-N bonds by C-B, C-N, or C-C bond gives rise to a considerable energy gain. In particular, the C$_2$-C$_3$ and C$_1$-C$_4$ and the C$_1$-C$_2$ carbon dimers at the $4|8$ (BB,NN) boundary resolve one and two high energetic bonds in the 4-atom ring and lower the formation energy by $\sim$1.5~eV and 3.1~eV, respectively. A pair of carbon substitutional defects in the 8-ring is less favorable. In contrast, there are no high energy B-B and N-N bounds at the $4|8$ (BN) boundary, thus the formation energies of boundary-carbon dimers are either comparable to or slightly higher than the far-site carbon dimer. Considering the example of $5|5|8$ boundaries, we always find high energetic bonds, thus we can identify favorable carbon pairs. For instance, the replacement of any of the B and N atoms at lattice sites 1, 3, 6, and 9 for the $5|5|8$(BN) grain boundary gives rise to a stable substitutional carbon dimer where at least one of the frustrated bonds\cite{Charlier1999} are relaxed. In addition to the high-energy-bond-relaxation mechanism, we observe an additional effect that influences the formation energy. The 5-ring environment with more N atoms is preferred. Similarly, sites 1 and 2 are preferred in $5|5|8$ with B-B or N-N bonds at the boundary.  In the case of $6|6$ grain boundaries, we identify only one dimer site at the junction, which is energetically favorable compared to the far-site dimer, see Fig.~\ref{fig:fig_GB}. Finally, the complex $5|7$ grain boundary also shows numerous favorable carbon dimer sites. The energetically most beneficial configurations are those that resolve either a B-B or N-N bound. The energy difference between the most and least favored site in $5|5|8$ and $5|7$ grain boundaries are comparable to the trends observed for the SW-C$_{\text{2}}$ (see Supplementary Fig.~S1) Note that there are also non-stoichiometric carbon dimer configurations when the carbon atoms substitute either an adjacent pair of B atoms or an adjacent pair of N atoms. Since iso-element pairs are unfavorable in BN, carbon substitutions at these sites could also be highly favorable, however, the formation energy would depend on the chemical potential of the involved elements.

Our results clearly show that there are preferred carbon dimer configurations at grain boundaries, where the most favorable ones are resolving the two high-energy B-B or N-N bonds. The formation energy of the most favorable carbon dimer configurations in our study is lower  by $\sim$3~eV than the bulk-like configuration. Taking into consideration the 2.2~eV formation energy (obtained with HSE(0.4) functional) of the carbon dimer in bulk hBN \cite{mackoit-sinkeviciene_carbon_2019}, the formation energy of the most favorable carbon dimers at the grain boundary becomes negative. This is only possible by lowering the formation energy of the grain boundary itself. Our results in agreement with Refs.~\cite{Li2012,Gomes2013} show that carbon atoms can further stabilize grain boundaries by replacing high-energy bonds.

\subsection{Kinetic stability and dynamical processes of SW-C defect formation}

In order to assess the stability of the SW-C defects, we study the height of the energy barrier between the substitutional carbon cluster and the corresponding SW-C configuration. In thermal equilibrium, the probability of the stable and the metastable configurations is determined by the formation energy difference of the configurations. At low temperatures, when $k_B T << E_b$, where $E_b$ is the reaction barrier between the two configurations, the atomic configuration may stuck in a higher energy configuration.

\begin{figure}
\begin{center}
    \includegraphics[width=0.5\textwidth]{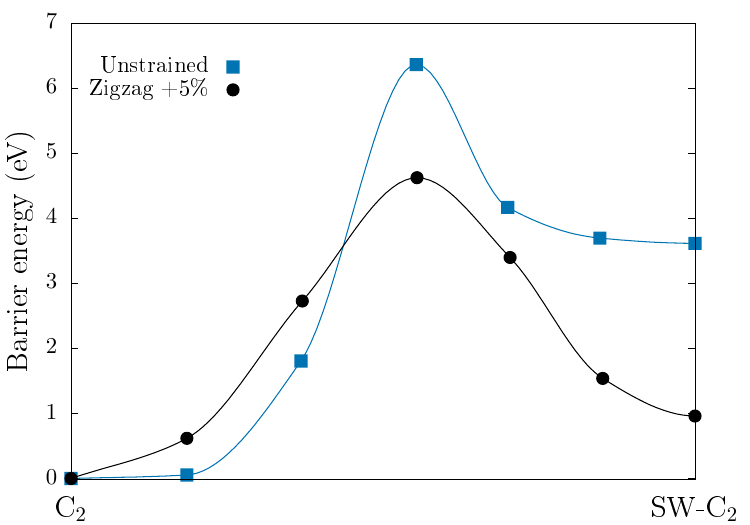}
\end{center}
	\caption{Barrier energy for transformation of substituted C$_{2}$ defect to SW-C$_{2}$.}
	\label{fig:fig_C2barrier}  
\end{figure}

Fig.~\ref{fig:fig_C2barrier} depicts the reaction pathway and corresponding energy for transforming a C$_2$ cluster into a SW-C$_2$ configuration. We considered both strained and unstrained hBN sheets. The energy barrier of the C$_2$~$\rightarrow$~SW-C$_2$ process is found to be 6.3~eV, while the barrier for the reversed process is 2.7~eV. Considering the intermediate steps of the transition, the highest energy point is associated with the 56$^{\circ}$ rotated and buckled configuration of the carbon dimer. We conclude that both the carbon clusters and the SW-C defect are kinetically stable under ambient temperatures, i.e.\    no transition is expected between the stable and metastable configurations.

These findings are not in favor of the formation of SW and SW-C defects. On the other hand, as demonstrated for graphene, there are alternative mechanisms that can give rise to SW and SW-C defects in hBN. The recombination of divacancy and a mobile carbon adatom can yield a SW-like defect in graphene \cite{Wu2013}, while a nitrogen-containing vacancy can lower the barrier to form SW-N in graphene\cite{Hou2015}. Here, the sliding motion of lattice-carbon atom facilitates the absorption of migrating atoms.
In addition, irradiation may transfer sufficient amount of energy to the lattice to form SW and SW-C configurations. Such transition may be more frequent for C clusters than for pristine hBN, due to the lower formation energy and lower energy barrier of the former. Taking into account the positive effect of tensile strain on the formation of SW-C defects, such defects may form in higher concentration  at strained areas under irradiation.

According to a recent theoretical work in Ref.~[\citen{weston_native_2018}], carbon and boron interlayer interstitial atoms are mobile at room temperature. Therefore, they either diffuse out from the layers or recombine with another defect in lattice. We study the energetics of various steps of mobile carbon atom-grain boundary recombination. As an example, we consider the case of $4|8$ (BB,NN) grain boundary and two  carbon adatoms.

\begin{figure}
\begin{center}
    \includegraphics[width=0.5\textwidth]{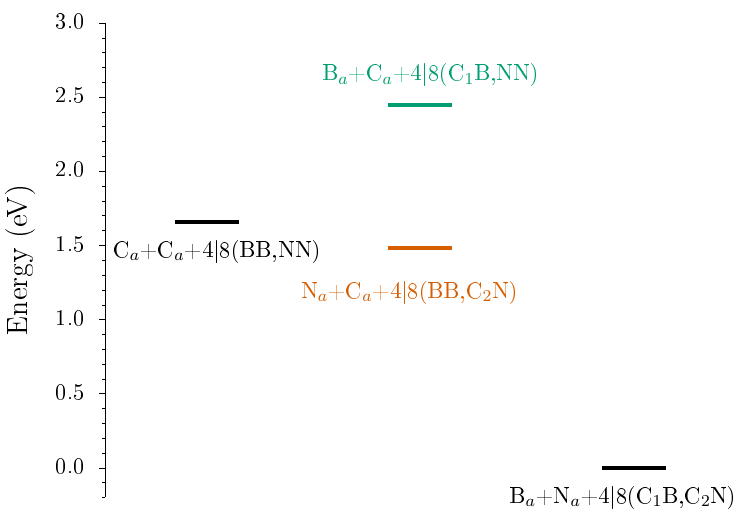}
\end{center}
 \caption{Energetics for carbon dimer formation at the  4$|$8(BB,NN) grain boundary. Colored lines with labels indicate the energy of different configurations of a system of a single-layer grain boundary and two adatoms. The number of atoms and electrons are the same in all configurations, thus the energies are directly comparable. C$_a$, B$_a$, and N$_a$ label different adatoms, while $4|8$(BB,NN), $4|8$(C$_1$B,NN), $4|8$(BB,C$_2$N), and $4|8$(C$_1$B,C$_2$N) label grain boundary configurations with no carbon contamination, one carbon atom substituting the boron atom at site 1, one carbon atom substituting the nitrogen atom at site 2, and two carbon atom substituting both the boron atom at site 1 and the nitrogen atom at site 2, respectively. The adatoms are placed far away from the grain boundary in all cases. We note that the energies may be lowered depending on the charge states\cite{PhysRevB.87.035404}.
 }
\label{fig:fig_pathway}  
\end{figure}

Fig.~\ref{fig:fig_pathway} shows the local minima of the configuration energy landscape of a $4|8$ (BB,NN) grain boundary with two carbon adatoms. Going from left to right, the carbon adatoms replace the boron and nitrogen atoms one by one and on the right reach an energetically favorable configuration where the two carbon atoms are incorporated into the hBN layer and a boron and a nitrogen atom are kicked out. Substituting the boron atom at the grain boundary is energetically unfavourable, however, the substitution of the nitrogen atom and creation of the nitrogen adatom is favoured. Importantly, substituting the adjacent B-N pair at the boundary with carbons, C$_1$-C$_2$ configurations, and creating both boron and nitrogen adatoms is energetically highly beneficial for the system. The substitution of B-N with C-C pair are frequently observed in TEM studies.\cite{Krivanek2010} Similarly, DFT study shows the absorption of carbon pair in 5$|$7 grain boundaries can aid the migration and bond rotation in defected areas\cite{Wang2015BN}. Therefore, we conclude that the incorporation of carbon atoms at the grain boundaries, even at the cost of displacing nitrogen and boron atoms is preferential, which may lead to the accumulation of carbon atoms and the formation of carbon-contaminated topological defects at the grain boundaries.

\subsection{Electronic structure and optical properties }

\begin{figure*}
\begin{center}
    \includegraphics[width=0.9\textwidth]{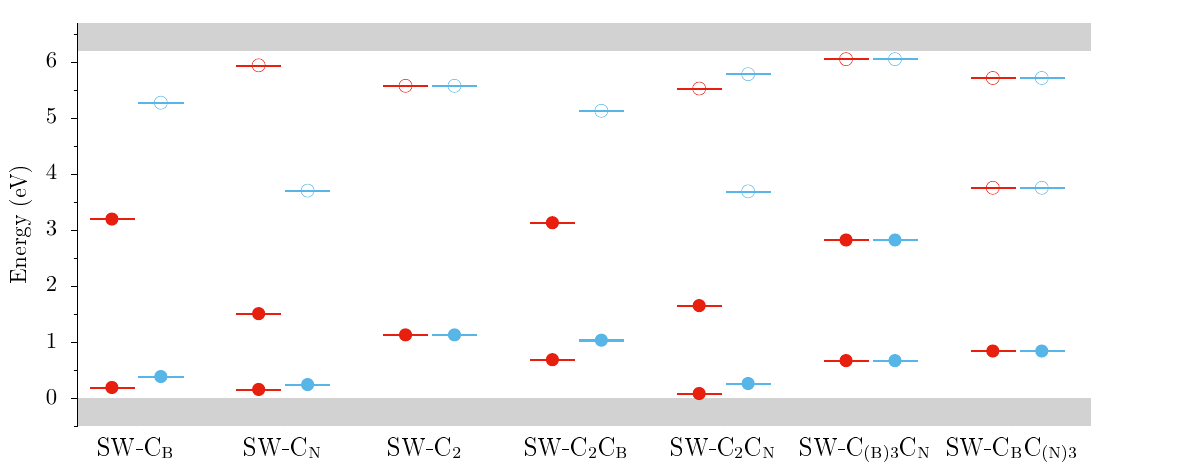}
\end{center}
	\caption{Kohn-Sham electronic structure of various neutral SW-C defects in hBN. Red and light blue lines depict spin-up and spin-down Kohn-Sham energy levels between the valence band and the conduction band (gray bars). The occupation of the defect states is indicated by filled and empty spheres. The electronic structure is obtained by using HSE(0.32) exchange-correlation functional.}
	\label{fig:fig_elec}  
\end{figure*}

\begin{figure*}
\begin{center}
    \includegraphics[width=0.95\textwidth]{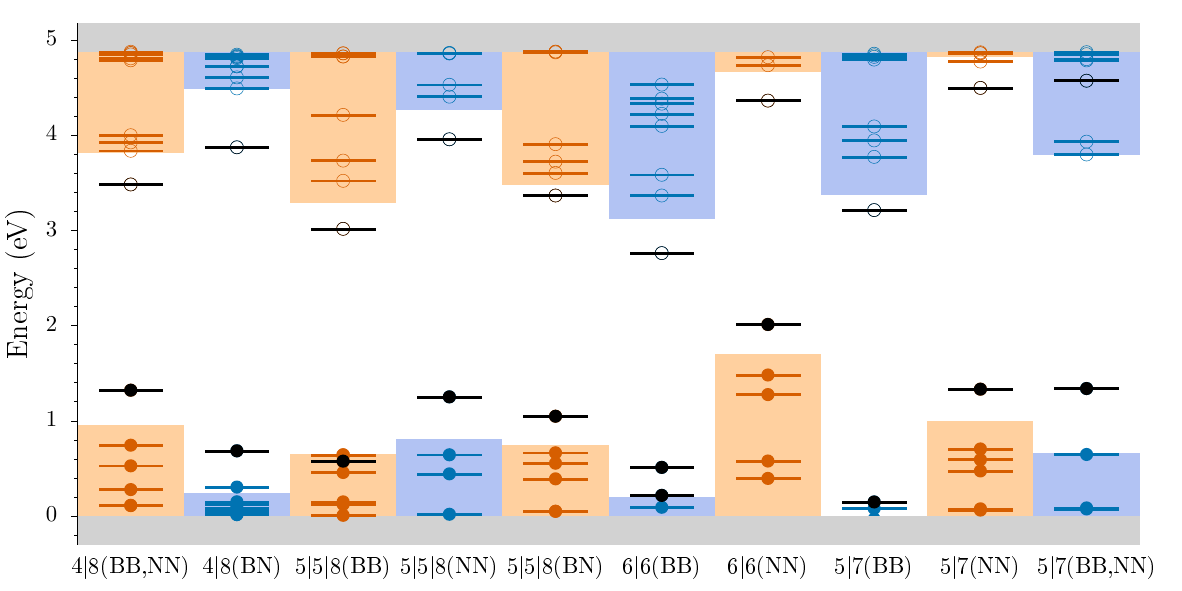}
\end{center}
	\caption{PBE Kohn-Sham electronic structure of different highly stable grain boundary carbon dimer configurations in the neutral charge state in hBN.  Colored lines with filled and empty circles indicate occupied and unoccupied Kohn-Sham energy levels in the bandgap of hBN. The valance band maximum and the conduction band minimum of pristine hBN are depicted by gray bands. Black lines denote electronic states localized on carbon atoms. The colored (amber and light blue) bars indicate the position of the grain boundary states when no carbon contamination is included in the model. The electronic structures are obtained by using the PBE exchange-correlation functional, which underestimates the energy gaps. Nevertheless, the PBE results are qualitatively valid. 
 }
	\label{fig:fig_elec_GB}  
\end{figure*}

In the previous sections, we analyzed the formation energy, stability, and alternatives of  formation of carbon-contaminated SW configurations and grain boundaries. The appearance of the SW-C defects and related configurations may play a role in defining the electrical, optical, and magnetic properties of hBN samples. Considering magnetism, we either find singlet, spin-0, or doublet, spin-1/2, ground state for the SW-C defects and carbon-containing grain boundaries depending on the parity of the total number of electrons, see Fig.~\ref{fig:fig_elec} and Table~\ref{tab:ZPL}. For carbon dimers in the SW-C and grain boundary configurations, we obtain a spin-singlet ground state due to the even number of electrons in the neutral charge state, see Figs.~\ref{fig:fig_elec} and \ref{fig:fig_elec_GB}. Since none of the studied configurations possess a high-spin ground state, we anticipate that these defects cannot implement spin qubits in hBN. On the other hand, we find both occupied and unoccupied defect states in the bandgap of hBN for the SW-C defects, see Fig.~\ref{fig:fig_elec}, as well as for the energetically favorable carbon-contaminated grain boundary configurations, see Fig.~\ref{fig:fig_elec_GB}. 

In the neutral charge state, most of the configurations exhibit occupied defect states in the lower half of the bandgap and unoccupied states in the upper half of the bandgap. Similar trends are observed in C-SW defects in hBN and nanotubes\cite{KIM201279,Anafcheh2013,Wang2016}. We anticipate from these observations that the neutral charge state of most of the SW-C and carbon dimer containing grain boundary defect configurations is stable when the Fermi energy is in the middle of the band gap of hBN. 

It is worth noting that the band gap of grain boundary models is significantly reduced compared to pristine hBN, due to the quasi-localized grain boundary states. The 4.8~eV PBE band gap of hBN reduces to 2.6-4.2~eV, in particular 4$|$8(BB,NN): 2.8~eV, 4$|$8(BN): 4.2~eV; $5|5|8$(BB):~2.6 eV; $5|5|8$(NN):~3.4 eV; $5|5|8$(BN): 2.7~eV; $6|6$(BB):~2.9 eV; $6|6$(NN):~2.9 eV; $5|7$(BB):~3.8 eV; $5|7$(NN):~3.9 eV; $5|7$(BB,NN):~3.1 eV, see colored shaded areas in Fig.~\ref{fig:fig_elec_GB}. Since both the conduction and the valence band edges are affected by the grain boundary states, grain boundaries can act as charge carrier traps. Furthermore, we notice in the calculations that the energy of the grain boundary states is often altered by the carbon dimer, which indicates strong hybridization between the states of the line and point defects.

In Figs.~\ref{fig:fig_elec} and \ref{fig:fig_elec_GB}, all the electronic structures exhibit at least one occupied as well as at least one unoccupied defect state in the band gap of either hBN or the considered grain boundary configuration. These results suggest that the SW-C configurations and carbon dimers at the grain boundary can give rise to color centers in the visible spectral range. In low defect concentrations, they may behave like single photon emitters in hBN.

\begin{table}[h]
\begin{center}
\caption{\label{tab:ZPL}  Ground state spin, ZPL energy, excitation lifetime of the considered SW-C defects in hBN. The ZPLs are obtained on HSE(0.32) level of theory.} 
 \begin{tabular}{c|c|c|c|c}
 \hline
Defect & Spin & ZPL (eV) & $\tau$ (ns) & DW  \\ 
 \hline  \hline
SW-C$_{\text{B}}$ & 1/2 & 1.96 & 44.8 &\\
SW-C$_{\text{N}}$ & 1/2 & 1.17 & 3.0 $\times 10^{3}$ &\\
SW-C$_{\text{2}}$ & 0 & 3.93  & 2.6 & 0.18\\
SW-C$_{\text{2}}$C$_{\text{B}}$ & 1/2 & 2.12 & 27.1 &\\
SW-C$_{\text{2}}$C$_{\text{N}}$ & 1/2 & 2.74 & 7.8 & \\
SW-C$_{\text{(B)3}}$C$_{\text{N}}$ & 0 & 2.62 & 15.1 & 0.09\\
SW-C$_{\text{B}}$C$_{\text{(N)3}}$ & 0 & 2.76 & 8.5 & 0.15\\
\hline
 \end{tabular}
 \end{center}
\end{table}

To further investigate this possibility, we first calculate the zero-phonon line (ZPL) energy of different SW-C configurations in hBN using HSE(0.32) functional, which are summarized in Table~\ref{tab:ZPL}. Similar results are obtained from TD-DFT calculations (see Methods and Supplementary Information for additional details). We find that the SW-C$_2$ defect emits at around 3.9~eV, while the most favorable four carbon-containing configurations are around 2.7~eV. The estimated error margin of the ZPL values is expected about 0.1-0.2~eV in our calculations. Thus, SW-C$_{\text{2}}$ and SW-C$_{\text{4}}$ defects emerge as alternative candidates for the UV and blue emitters.\cite{mackoit-sinkeviciene_carbon_2019,maciaszek_blue_2024}

In order to estimate the brightness of the SW-C color centers, we calculate the radiative lifetime of the SW-C$_{\text{2}}$, SW-C$_{\text{(B)3}}$C$_{\text{N}}$, and SW-C$_{\text{(B)}}$C$_{\text{(N)3}}$ configurations. We respectively obtain 2.6 ns, 15.1 ns, and 8.5 ns for the optical lifetime of these defects. Due to the relatively large ZPL and the presumably absent alternative spin shelving states, we expect high quantum efficiency and bright visible emission for these defects. 
\begin{figure*}[h]
\begin{center}
\includegraphics[width=0.49\textwidth]{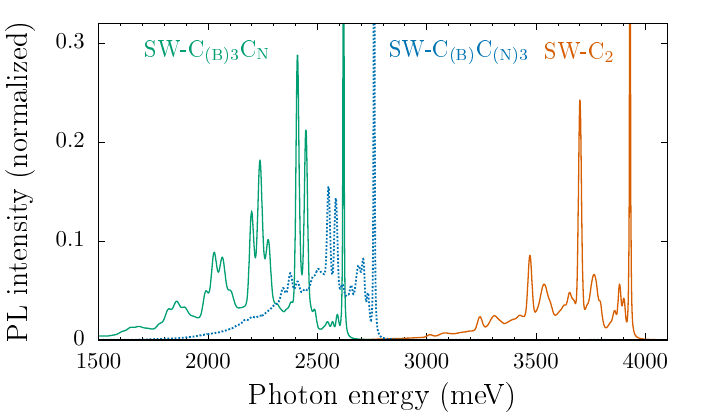}
\includegraphics[width=0.49\textwidth]{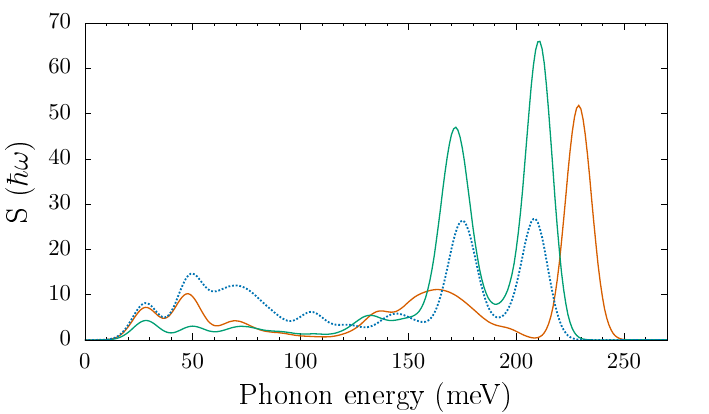}
	\caption{ Calculated photoluminescence spectra and phonon energy spectral function for select SW-C defects. The upper panel denotes the normalized PL spectra for SW-C$_{\text{2}}$ (orange), SW-C$_{\text{(B)3}}$C$_{\text{N}}$ (green) and SW-C$_{\text{(B)}}$C$_{\text{(N)3}}$ (blue) defects while the lower panel shows the spectral function S for phonon modes contributing to the PL spectra. The unequal carbon bonds of SW-C$_{\text{(B)3}}$C$_{\text{N}}$ result in two phonon sidebands at 171 and 210 meV.}
	\label{fig:fig_PL1}  
 \end{center}
\end{figure*}
Accordingly, we determine their PL spectra, depicted in Fig.~\ref{fig:fig_PL1}. As can be seen, the C-C phonon modes are similar in nature to the carbon substitutional defects albeit with energy shifts corresponding to the altered bond strengths in SW configuration. The Debye-Waller factors for SW-C$_{\text{2}}$ and SW-C$_{\text{(B)}}$C$_{\text{(N)3}}$ are found to be comparable to the values predicted for hBN SW and carbon substitutional defects.\cite{mackoit-sinkeviciene_carbon_2019,hamdi_stonewales_2020,Benedek2023}   

\begin{table}[!h]
\begin{center}
\caption{\label{tab:ZPL-GB} ZPL energy, excitation lifetime of the highly stable carbon dimers at $4|8$ grain boundaries in hBN. The ZPLs are obtained on HSE(0.32) level of theory.} 
 \begin{tabular}{l|c|c|c|c}
 \hline
Grain boundary & Configuration & ZPL (eV) &$\tau$ (ns) & DW  \\ 
 \hline  \hline
$4|8$(BB,NN) & C$_1$-C$_2$ & 2.72 & 16.6 & 0.02\\
$4|8$(BN) & C$_3$-C$_4$ & 3.84 & 2.0 & 0.24\\
\hline
 \end{tabular}
 \end{center}
\end{table}

\begin{figure}[!]
\includegraphics[width=0.49\textwidth]{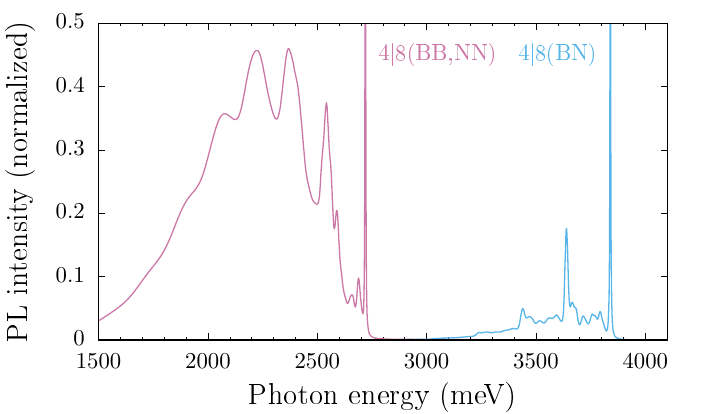}
\includegraphics[width=0.49\textwidth]{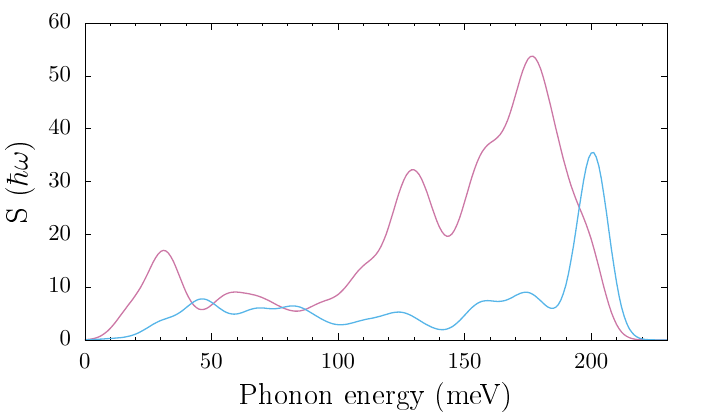}
	\caption{ Calcauted photoluminescence spectra and phonon energy spectral function for carbon dimers at grain boundaries $4|8$(BB,NN) (pink) and $4|8$(BN) (blue).}
	\label{fig:fig_PL2}  
\end{figure}

We also explore the optical properties of grain boundaries which largely depend on the local bonding at the boundary.
Owing to computational demand, we consider only a few  highly favorable carbon dimer-containing grain boundary configurations as illustration, see Table~\ref{tab:ZPL-GB}.
We find that the studied configurations emit in a similar spectral range as the SW-C defects.
The corresponding radiative lifetime for the most stable carbon dimers in $4|8$(BB,NN) and $4|8$(BN) is 16.6 ns and 2.0 ns respectively.
As shown in Fig.~\ref{fig:fig_PL2} and in Table~\ref{tab:ZPL-GB}, the C$_1$-C$_2$ carbon dimer in $4|8$(BB,NN) grain boundary emits in the visible range with a predominant phonon contribution to the PL spectra. In contrast, the UV active C$_3$-C$_4$ defect hosted  in the $4|8$(BN) grain boundary  exhibits a significant Debye-Waller factor (0.24). Therefore, we expect other grain boundaries to host bright color centers as well. Point defects may act as a recombination center for excitons and separated electron and hole pairs captured by the quasi-localized states of the grain boundary. Realizing charge carrier injection in such structures may lead to electroluminescence.\cite{Rong2015,Kim2020}


From the computed optical properties of the SW-C defects and carbon-contaminated grain boundaries, we conclude that some of these defects are particularly interesting for quantum photonic applications due to their short lifetime and dominating ZPL emission. In addition, the optical properties of the SW-C defect resemble the carbon-related visible emitters.\cite{Mendelson2019,zhigulin_photophysics_2023}

\section{Discussion}

In this article, we demonstrated that carbon contamination can play an important role in stabilizing topological defects, such as Stone-Wales defects and grain boundaries in hBN. Carbon Stone-Wales may appear in increased concentration in strained areas of irradiated hBN, while carbon impurities may accumulate at grain boundaries and form a series of highly stable carbon-containing defect structures. We also showed that the studied defects can give rise to bright color centers in hBN emitting in the visible spectral range with desirable optical properties. Based on the similarity of the optical properties, the SW-C defect may serve as a working model for the not-yet-identified carbon-related visible emitters in hBN. While these defects may be readily observed in as-grown samples, imparting sufficient energy to facilitate bond rotation via irradiation may increase the number of SW-C defects. Indeed, recent experiments\cite{kumar_polarization_2024} used such an aftergrowth treatment and reported on an increase of visible emitters. 
To verify our models, the alignment of the emission dipole and the crystal lattices may be compared. Due to the 90$^\circ$ rotation of a carbon-related bond in the SW-C defects, the emission dipole 
. Consequently, correlating crystallographical directions with the emission dipole, one may observe 30$^\circ$, 90$^\circ$, and 150$^\circ$ emission dipole orientations compared to one of the inplane axes for simple SW-C configurations.  In a very recent measurement, see Fig.~2e in Ref.~[\citen{kumar_polarization_2024}], a similar pattern has been reported. However, more measurements are needed to make conclusions here.

The defects studied in this article not only broaden our understanding of hBN's defect landscape but also point to a potential avenue for identifying previously unknown hBN color centers. Ultimately, our findings may pave the way toward the controlled fabrication of visible emitters in hBN, opening new possibilities for applications in quantum technologies and optoelectronics.

{\section{Methods}}

Density functional theory (DFT) calculations for single-layer hBN models are performed using the Vienna Ab-initio Simulation Package (VASP)\cite{VASP2}, which utilizes the projector-augmented wave method\cite{PAW} and a plane-wave basis set. The plane-wave and kinetic energy cutoff are set to 450 and 900~eV, respectively. Exchange-correlation effects of the many-electron system are described by the generalized gradient approximation of Perdew, Burke, and Ernzerhof (PBE)\cite{PBE} and also the screened hybrid functional of Heyd-Scuseria-Ernzerhof (HSE06)\cite{HSE06} with a modified exact exchange fraction of 0.32 and screening parameter of 0.2~\AA $^{-1}$. The PBE functional is used in the study of defect formation energies and grain boundary electronic structures, while the HSE06 functional is used to obtain ZPL energies and the electronic structure of point defects. The van der Waals interaction is included using the Grimme-D3 method\cite{Grimme2010}. 

In our study, we consider a 128-atom hexagonal supercell for the SW-C defect calculations including a vacuum of 18~\AA\  between the periodic hBN layers. The grain boundaries are modeled in orthorhombic supercells of dimensions 18-20~\AA\  along with the grain boundary and varying lengths of 30-80~\AA\ in the perpendicular direction. Following the coincidence site lattice theory,\cite{PhysRevB.84.165423} the periodicity of supercell is retained by including a single, parallel, or antiparallel grain boundary. The electronic structure of carbon dimer containing grain boundaries is also verified in hydrogen-terminated ribbon models for selected cases. The reciprocal space is sampled with $\Gamma$-point. Atomic positions are relaxed until the forces are less than 0.01 eV/\AA. The bond rotation barrier energies are calculated with PBE functional using the climbing image nudged elastic band method~\cite{Henkelman2000a,Henkelman2000b}. 

The excited state properties are obtained using a combination of electronic and vibrational calculation setups. The ZPL energy is calculated as the energy difference between ground and excited states, where the excited state configuration is found using $\Delta$-SCF method\cite{RevModPhys.61.689}. In case of defects with singlet spin configuration, we estimate the energy of excited singlet state as 2$E(M)$-$E(T)$ where $E(M)$ refers to the energy of default mixed-spin excited state and $E(T)$ is the energy of the relaxed triplet state\cite{mackoit-sinkeviciene_carbon_2019}. The photoluminescence spectra and the Debye-Waller factor are calculated with the PyPhotonics package\cite{Tawfik2022}. Here, the phonon modes are calculated from Phonopy code\cite{Togo2015,phonopy-phono3py-JPSJ} using the finite-displacement method. The grain boundary supercell dimensions are reduced to manage the high computational cost of phonon calculations.

The radiative lifetime is calculated as\cite{stoneham2001theory}
\begin{equation} 
    \begin{split}
       \tau = 
        \frac{3\pi \epsilon_{0} c^{3} {\hbar}^4}{nE_\text{ZPL}^{3}\mu^{2}}\text{,}
    \end{split}
\end{equation}
where $\epsilon_{0}$ is the vacuum permittivity, $c$ is the speed of light, $n$ is the refractive index of hBN and $\mu$ is the transition dipole moment of the electronic states corresponding to the ZPL. The refractive indices are taken from an ellipsometry study on thin film hBN\cite{Schubert1997}.

To verify the ZPL energies calculated with $\Delta$-SCF method, we carry out time-dependent DFT (TD-DFT) calculation for hBN flake. The vertical and ZPL energy calculations for defects with singlet spin configuration are performed using the ORCA package\cite{neese2022software}. A combination of PBE0\cite{PBE0}, HSE06, CAM-B3LYP\cite{CAMB3LYP} functionals and def2-SVP\cite{weigend2005balanced}, def2-TZVP, aug-cc-pVDZ\cite{augccpvdz} basis sets for 58 to 158-atom flakes are considered. The first excited singlet state is identified by invoking five roots in TD-DFT calculation, followed by geometry optimization.

\section*{ Data availability}

The main data supporting the findings of this study are available within the paper and its Supplementary Information. Further numerical data are available from the authors upon reasonable request.

\section*{Author contributions}

R.B. designed the project and carried out the first-principles calculations with inputs from A.G. The manuscript was written by V.I., R.B., and G.B. with inputs from all co-authors. The work was supervised by V.I. and G.B.

\section*{Competing interests} 

The authors declare no competing interests.
 
\section*{Acknowledgments} 

This research was supported by the National Research, Development, and Innovation Office of Hungary within the Quantum Information National Laboratory of Hungary (Grant No. 2022-2.1.1-NL-2022-00004) and within grants FK 135496 and FK 145395.
V.I. and I.A.A. also acknowledge the support from the Knut and Alice Wallenberg Foundation through WBSQD2 project (Grant No.\ 2018.0071). 
The computations were
enabled by resources provided by the National Academic
Infrastructure for Supercomputing in Sweden (NAISS)
and the Swedish National Infrastructure for Computing
(SNIC) at NSC, partially funded by the Swedish Research
Council through grant agreements no. 2022-06725
and no. 2018-05973.
We acknowledge KIF\"U for awarding us access to computational resources based in Hungary.

\end{document}